\documentclass[10pt,letterpaper]{article}
\usepackage{opex3}
\usepackage{amssymb}
\usepackage{graphicx}
\usepackage{verbatim}
\usepackage{eucal}
\usepackage{cite}

\newcommand{\bra}[1]{\ensuremath{\left\langle #1 \right|}}
\newcommand{\ket}[1]{\ensuremath{\left| #1 \right\rangle}}

\begin{document}

\title{A wavelength-tunable fiber-coupled source of narrowband entangled photons}

\author{Alessandro Fedrizzi$^1$, Thomas Herbst$^1$, Andreas Poppe$^2$, Thomas Jennewein$^1$ and Anton Zeilinger$^{1,2}$}
\address{$^1$Institute for Quantum Optics and Quantum Information, Austrian Academy of Sciences,
Boltzmanngasse 3, 1090 Wien, Austria
 \\$^2$Quantum Optics, Quantum Nanophysics and Quantum Information, Faculty of Physics, University of Vienna,
    Boltzmanngasse 5, 1090 Vienna, Austria}
\email{zeilinger-office@univie.ac.at}


\begin{abstract}
We demonstrate a wavelength-tunable, fiber-coupled source of
polarization-entangled photons with extremely high spectral
brightness and quality of entanglement. Using a $25~$mm PPKTP
crystal inside a polarization Sagnac interferometer we detect a
spectral brightness of $273000$~pairs~(s~mW~nm)$^{-1}$, a factor of
28 better than comparable previous sources while state tomography
showed the two-photon state to have a tangle of $T=0.987$. This
improvement was achieved by use of a long crystal, careful selection
of focusing parameters and single-mode fiber coupling. We
demonstrate that, due to the particular geometry of the setup, the
signal and idler wavelengths can be tuned over a wide range without
loss of entanglement.
\end{abstract}

\ocis{(270.0270) Quantum Optics; (190.4410) Nonlinear Optics, parametric processes;} 


\section{Introduction}
Sources for photonic entanglement form an integral part of quantum
optics experiments and quantum information protocols like quantum
key distribution \cite{ekert1991,jennewein00}, quantum teleportation
\cite{bennett93,bouwmeester97} and quantum computing
\cite{knill2001}. To date the most successful method of creating
polarization entangled photon pairs is spontaneous parametric
downconversion (SPDC) in nonlinear crystals. In the first practical
and efficient scheme, orthogonally polarized photon pairs were
created in a Type-II BBO bulk crystal and emitted along intersecting
cones \cite{kwiat95}. Since then nonlinear optics has advanced;
Periodic poling of nonlinear crystals considerably relaxes the
phasematching conditions for SPDC and allows to fully exploit the
material's nonlinear properties. These crystals are best employed in
collinear configuration, where a much bigger fraction of the created
photons can be entangled than in the conelike geometry of
\cite{kwiat95}, thus leading to downconversion sources of higher
spectral brightness. This increase comes at the cost of a new
problem though, the need to spatially separate the collinear
downconversion modes.

When the downconverted photons are created at substantially
different wavelengths, they can be separated by dichroic mirrors,
and several demonstrations of this have reported high spectral
brightness \cite{pelton04,konig05,hubel07}. For many applications,
however, wavelength-degenerate photons are preferable. The first
attempt to build a source based on periodically poled KTiOPO$_4$
(PPKTP) and collinear beam propagation used a simple beamsplitter to
separate the output modes, at the cost of an unwanted $50\%$ loss in
the coincidences \cite{kuklewicz04}. A more effective method,
originally proposed in \cite{kwiat94}, is to interferometrically
combine the outputs of two downconverters on a polarizing
beamsplitter. The authors of \cite{fiorentino04} implemented this
idea, using one bidirectionally pumped PPKTP crystal in a folded
Mach-Zehnder interferometer. The spectral brightness of this source
reached $4000$~pairs(s mW nm)$^{-1}$, but the Mach-Zehnder
interferometer was sensitive to vibrations and even with active
phase stabilization the visibility of the setup was limited to
$90\%$. To avoid the need to actively stabilize their setup, the
authors of \cite{shi04} combined the downconversion pairs created in
a type-I BBO crystal on a symmetric beamsplitter in an intrinsically
phase-stable Sagnac interferometer. This configuration proved to be
stable, but the achieved count rates and visibilities did not come
close to the configurations realized before. Finally, in a recently
published work Kim et al. \cite{kim06} ingeniously combined the
advantageous features of \cite{fiorentino04} and \cite{shi04},
bidirectionally pumping a type-II PPKTP crystal in a polarization
Sagnac interferometer. The authors reported a detected photon-pair
yield of $5000$~(s~mW~nm)$^{-1}$ at a $96.8\%$ visibility. In an
update \cite{wong06}, these numbers were significantly improved to
$9800$~(s~mW~nm)$^{-1}$ at $~98\%$ visibility.

During this time, there has also been significant improvement in the
development of correlated photon-pair sources based on four-wave
mixing in single-mode fibers \cite{fan07, liang07} and SPDC in
nonlinear crystal waveguides \cite{spillane07}. The strong spatial
confinement of the pump beam in nonlinear waveguides leads to higher
effective material nonlinearities and thus the generation of photon
pairs of much higher spectral brightness than in bulk periodically
poled crystals. Yet, to date no high quality polarization
entanglement has been demonstrated in these schemes.

In this paper we demonstrate a fiber-coupled source of
polarization-entangled photons, based upon the scheme by Kim et al.
\cite{kim06}. We show that the wavelength of the downconversion
photons can be tuned without loss of visibility over a range of
$\pm26$~nm around degeneracy merely by changing the temperature of
the nonlinear crystal. Wavelength tunability of entangled photons
has been claimed before \cite{konig05, hubel07} but was not
demonstrated to the extent achieved here. The high purity of the
entangled photon state in our setup enables further tests on
fundamental concepts of physics for example on the violation of
local or non-local realism \cite{groeblacher07, paterek07}, which
place increasingly stringent lower bounds on the required degree of
entanglement. The high spectral brightness of the source makes it
perfectly suitable for free-space quantum communication experiments
\cite{resch05,peng05,ursin07}. Given that the range of wavelengths
accessible to this source covers important atomic transitions in
rubidium or caesium, it is also a promising candidate for
atom-photon coupling experiments, e.g. photonic memories for quantum
repeaters \cite{matsukevich04,julsgaard04}.

The paper is organized as follows: in section \ref{sec:spdc} we
briefly review spontaneous parametric downconversion in periodically
poled crystals. In section \ref{sec:focussing} we discuss the
influence of the crystal length on the performance of a
downconversion setup as well as the optimal focusing conditions for
type-II PPKTP crystals of varying length. In section \ref{sec:setup}
we describe the source of polarization-entangled photons and in
section \ref{sec:results} we present our results on obtained
spectral brightness, quality of entanglement and wavelength
tunability.

\section{Downconversion in periodically poled crystals}
\label{sec:spdc} Parametric downconversion in a nonlinear crystal
can be described by the spontaneous splitting of a pump photon
(\textit{p}) into a pair of photons (usually called \textit{signal
s} and \textit{idler i}) according to conservation of photon energy
$\omega_p=\omega_s+\omega_i$ and momentum. In periodically poled
crystals, the effective nonlinearity of the medium is periodically
inverted by the application of an electric field with alternating
directions during the crystal fabrication process. In contrast to
birefringent phasematching in bulk crystals the quasi-phasematching
conditions now involve an additional term which depends on the
crystal-poling period $\Lambda$:
\begin{equation}
\textbf{k}_p(\lambda_p,n_p(\lambda_p,T))=\textbf{k}_s(\lambda_s,n_s(\lambda_s,T))+\textbf{k}_i(\lambda_i,n_i(\lambda_i,T))+\frac{2\pi}{\Lambda(T)}
\label{eq:momentum}
\end{equation}
where $\textbf{k}_{p,s,i}$ are the wavevectors of the pump, signal
and idler photons, and $n_{p,s,i}$ the wavelength and
temperature-dependent refractive indices of the crystal for the
respective wave fields. Assuming a fixed pump wavelength, Eq.
\ref{eq:momentum} shows that the $\textbf{k}$ vectors of the signal
and idler photons are temperature dependent, in particular, for a
fixed emission angle, wavelength-degenerate photons will only be
created at one certain temperature $T$. Quasi-phasematching in
periodically poled crystals allows for almost arbitrary
phasematching angles and wavelengths in comparison to the nontrivial
conditions in bulk nonlinear crystals, where only birefringent and
angle phasematching is possible. In particular, periodically poled
crystals can be tailored such that $\textbf{k}_{p,s,i}$ are parallel
to one of the crystallographic axes $X,Y$ or $Z$. For a type-II SPDC
process this means that the orthogonal signal and idler beams do not
experience transversal walkoff, an effect which severely limits the
maximum useful length of birefringent nonlinear crystals to less
than $10$~mm in typical downconversion setups. In section
\ref{sec:focussing} we compare photon-pair yields coupled to
single-mode optical fibers and the corresponding single-photon
bandwidths of downconversion photons created in periodically poled
crystals of various lengths. Furthermore we find the optimal
focusing conditions for each crystal resulting in unprecedented
photon yields.

\section{Optimization of beam focusing and crystal properties} \label{sec:focussing}
In downconversion experiments it is well known that focusing the
pump beam drastically improves count rates and collection
efficiencies into single-mode fibers. The quest for the optimal
focusing parameters has been the subject of theoretical and
experimental investigations
\cite{kurtsiefer01,bovino03,ljunggren05}. We followed the lines of
Ljunggren et al. \cite{ljunggren05}, who numerically calculated
optimal focusing conditions for downconversion in PPKTP crystals and
made predictions for the expected dependence of count rates and
photon bandwidths on crystal length. The two key parameters for an
entangled photon source which can be maximized via optimal focusing
are the detected coincidence rate $R_c$ and the
coincidence-to-single-photon (coupling) ratio
$\eta_c=R_c/\sqrt{R_iR_s}$, with $R_i$ and $R_s$ being the single
photon count rates. As pointed out in \cite{ljunggren05} the
focusing conditions for these parameters are not generally the same,
as the number of single photons and photon pairs coupled into fibers
do not solely depend on the total number of pairs created but also
on the mode quality factor $M^2$ of the downconversion and its
overlap with the fiber modes.

Here, instead of measuring $M^2$ factors as done in
\cite{ljunggren05} we directly compared single photon and
coincidence count rates for four ($L=10, 15, 20, 25$~mm) X-cut
flux-grown PPKTP crystals manufactured by \textit{Raicol}. The
crystals had a grating period of $10$~$\mu$m and were
quasi-phasematched for collinear degenerate downconversion at
$\lambda_p=405$~nm and $49.2\symbol{23}C$. The $405$~nm laser diode
used in our setup had limited power, therefore we chose to find the
optimal focusing conditions for a maximal coincidence rate
$R_{c~\rm{max}}$.

\begin{figure}[htbp]
\begin{center}
$\begin{array}{c@{\hspace{0in}}c}
\includegraphics[width=6.cm]{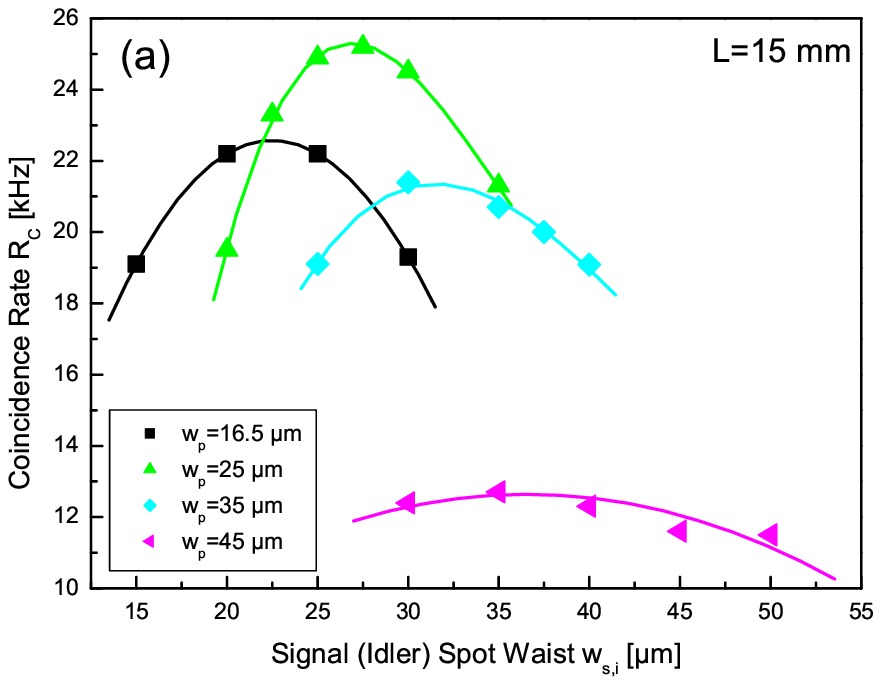} &
\includegraphics[width=6.cm]{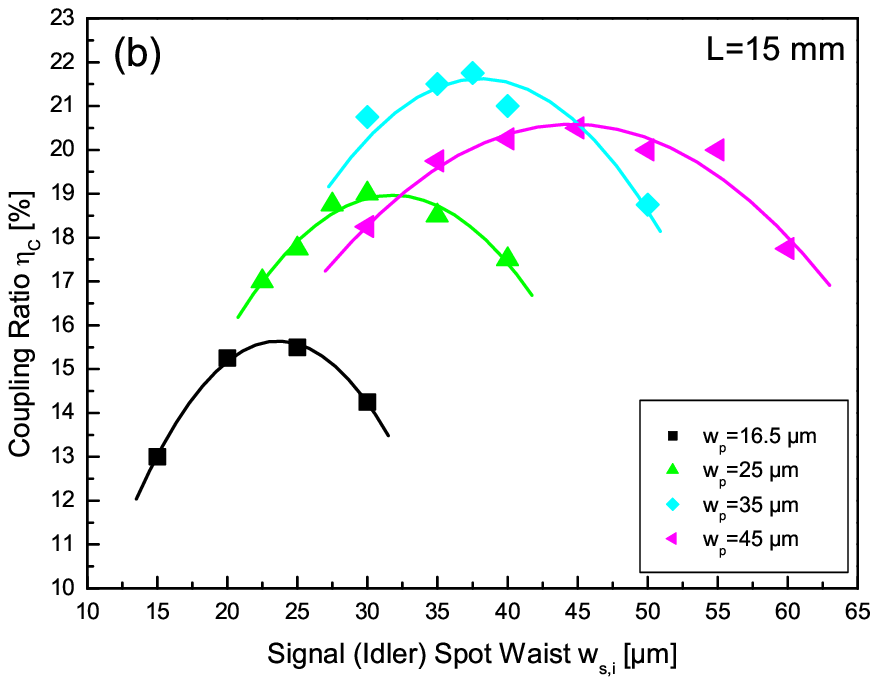} \\
\includegraphics[width=6.cm]{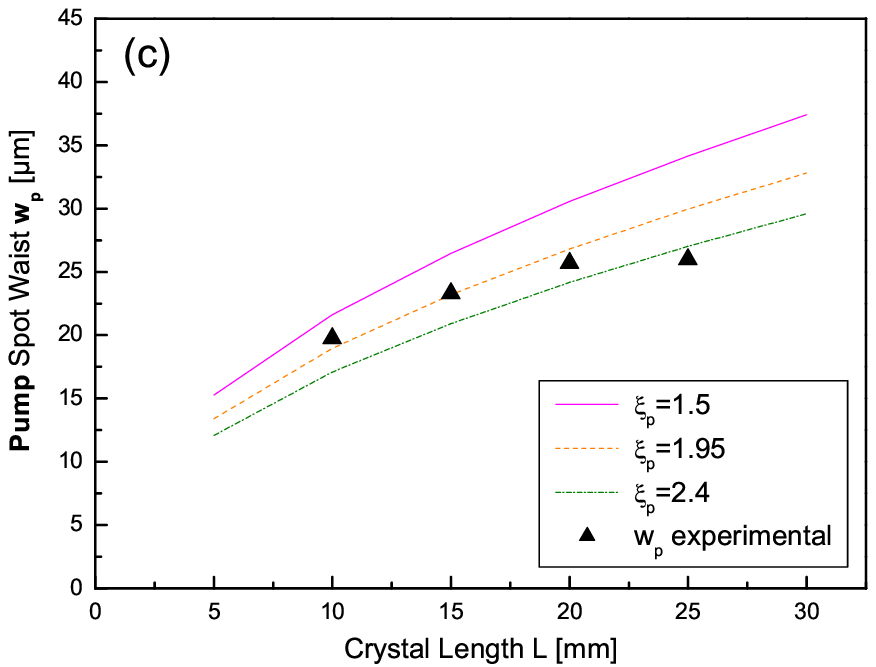} &
\includegraphics[width=6.cm]{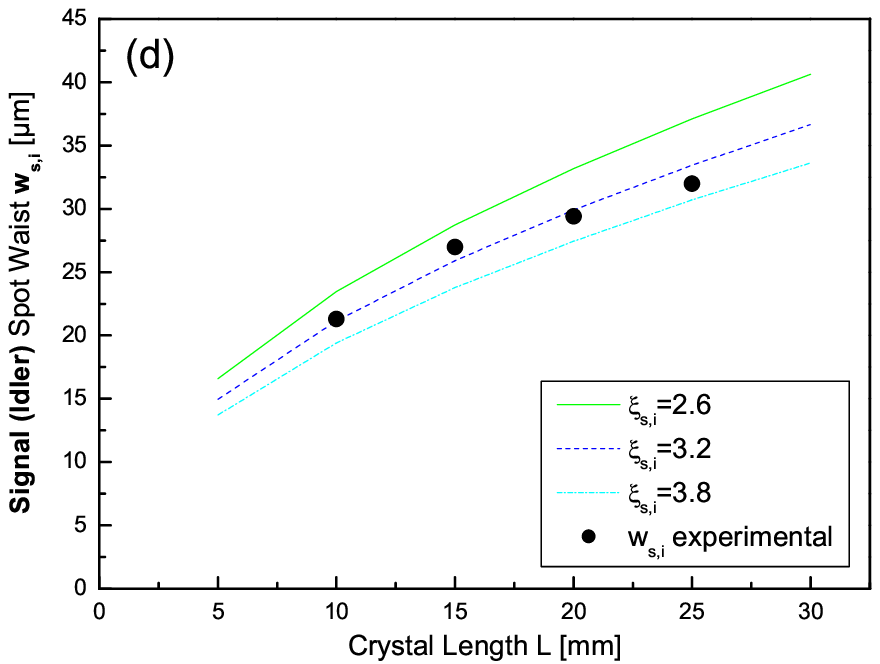}
\end{array}$
\end{center}
\caption{Measured coincidence count rates $R_c$ (a) and coupling
ratios $\eta_c$ (b) for a $L=15~$mm crystal for a series of focusing
conditions in a simple test setup (described in the text).  From
these measurements we determined pump (c) and signal (idler) (d)
spot waist sizes $w_p$ and $w_{s,i}$ for maximal photon-pair count
rate $R_{c~\rm{max}}$ and crystals of length $10, 15, 20$ and
$25$~mm. The detected count rates were repeatable to within $5\%$.
The lines drawn for three constant values of the dimensionless
parameters $\xi_p$ and $\xi_{s,i}$ ($\xi=L/z_r$, where $z_r$ is the
Rayleigh range) show the supposed independence of $\xi_{p~\rm{opt}}$
and $\xi_{s,i~\rm{opt}}$ of $L$. While this behavior is certainly
given for $\xi_{s,i~\rm{opt}}$ (black dots), it was not observed for
$\xi_{p~\rm{opt}}$. (black triangles).} \label{fig:focussing}
\end{figure}

We pumped a given crystal of length $L$ with spot waist sizes $w_p$
from $15-55~\mu m$, coupling the downconversion into a single-mode
fiber attached to a fiber beamsplitter and two detectors. For each
$w_p$ we varied mode waist sizes $w_{s,i}$ between $(15\mu
m<w_{s,i}<55\mu m)$ and monitored $R_{s,i}$ and $R_c$. The obtained
data, exemplary shown for the $L=15$~mm crystal in figures
\ref{fig:focussing}(a) and (b), was numerically fitted with
polynomial functions. From these fits, for each $w_p$ we obtained
one optimal $w_{s,i~\rm{opt}}$ and, in consequence one optimal set
$(w_{p}, w_{s,i})_{\rm{opt}}$ for each crystal length $L$.

The results in Fig. \ref{fig:focussing} (c) and (d) show the spot
waist sizes for maximal photon-pair yield for the pump beam and the
signal (idler) modes for different crystal lengths. For crystals
from $10-25$~mm the optimal focusing conditions for $R_{c~\rm{max}}$
are $(20\mu m\lesssim w_{p~\rm{opt}}\lesssim26\mu m)$ (Fig.
\ref{fig:focussing}(c)) and $(21\mu m\lesssim
w_{s,i~\rm{opt}}\lesssim32\mu m)$ (Fig. \ref{fig:focussing}(d)).
Using the dimensionless parameters $\xi_p=L/z_{r~p}$ and
$\xi_{s,i}=L/z_{r~s,i}$, where $z_{r~p,s,i}$ are the Rayleigh ranges
of the pump and the signal (idler) modes in the crystal, this
corresponds to $(1.8\lesssim\xi_{p~\rm{opt}}\lesssim2.6)$ and
$\xi_{s,i~\rm{opt}}\thicksim3.2$. We found that for the signal and
idler modes the focusing parameter $\xi_{s,i~\rm{opt}}$ is
independent of $L$ (cf. \cite{ljunggren05}), but that the pump focus
parameter $\xi_{p~\rm{opt}}$ rises with L, which might be due to
increasing contribution of grating defects in longer crystals.

For the $L=15$~mm crystal we could also deduce the focusing
conditions for a maximal coupling ratio
$\eta_{c~\rm{max}}=max~\eta_c(\xi_{p},\xi_{s,i})$. The parameters
$\xi_{p}$ and $\xi_{s,i}$ differ significantly from those for
$R_{c~\rm{max}}$ with $\xi_p\thicksim0.7$ and
$\xi_{s,i}\thicksim1.5$. The reason is presumably that for tighter
confinement of the pump field within the crystal, a higher total
number of pairs is generated, which results in a higher $R_c$. Due
to the stronger focusing these pairs will however be created in
spatial longitudinal modes of increasingly higher order (cf.
\cite{ljunggren05}). This leads to a decrease of the overlap between
the coupling fiber modes and the downconversion mode and hence to a
lower coupling ratio $\eta_c$.

Consider the properties of the power spectrum of the emitted
two-photon state, obtained by using first-order perturbation theory
\cite{rubin94}:
\begin{equation}
I\propto sinc^2(\frac{L}{2}\Delta k) \label{eq:sinc}
\end{equation}

where $\Delta k$ is the phase mismatch,
$k_p-k_i-k_s-\frac{2\pi}{\Lambda}$. The authors of
\cite{ljunggren05} conclude that the flux of downconversion photons
$R_c$ coupled into single-mode optical fibers in collinear
configurations with optimal focusing relates to the crystal length
$L$ as
\begin{equation}
R_c\sim\sqrt{L}.\label{eq:ratedependence}
\end{equation}
The $sinc^2$ function in Eq. \ref{eq:sinc} drops to a value of $0.5$
for $\frac{L}{2}\Delta k=1.39$. By writing $\Delta k$ as a function
of $\lambda_i$, expanding $\Delta k$ in a Taylor series around
$\lambda_{i}=810$~nm and neglecting higher order terms, we arrive at
an expression for the single-mode bandwidths of the signal and idler
photons:
\begin{equation}
\Delta\lambda_{s,i FWHM}=\frac{5.52x10^{-3}}{L}~\rm{nm}
\label{eq:bandwidth}
\end{equation}
The spectral brightness $B$ of the produced downconversion light, a
value for the comparison of different downconversion setups, given
by the number of produced photon pairs per second, per milliwatt of
pump power and per nm bandwidth, therefore scales as $B\sim
L\sqrt{L}$.
\begin{figure}[hbtp]
\centering\includegraphics[width=9cm]{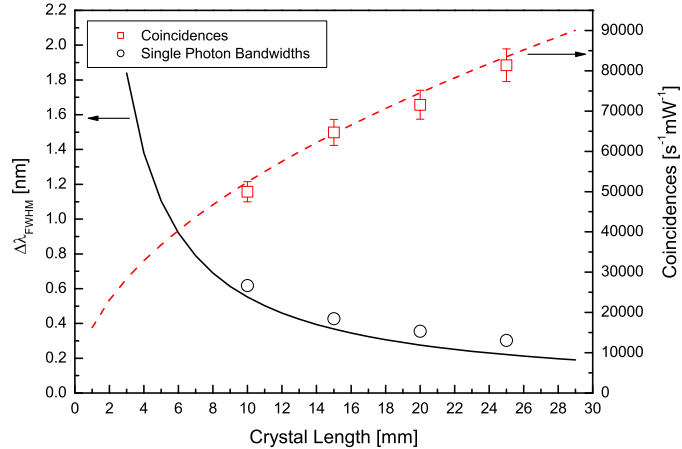}
\caption{Measured spectral bandwidths (FWHM) of downconversion
photons and photon-pair yields for PPKTP crystals of various
lengths. The single photon bandwidth (black circles) was determined
via interference in a single photon Michelson interferometer and is
shown compared to theoretic values calculated using Eq.
\ref{eq:bandwidth} (black line). The measured coincidence rates (red
squares) show the expected square root dependence on $L$ (Eq.
\ref{eq:ratedependence}).}\label{fig:lscaling}
\end{figure}

The experimental results on the photon-pair yield $R_c$ and the
bandwidth $\Delta\lambda_{s,i}$ of the downconverted photons for
different crystal lengths are shown in Fig. \ref{fig:lscaling}. The
maximum photon-pair yields $R_c$ resulted from the evaluation of the
optimal focusing conditions. A least-square fit of the corresponding
data points to Eq. \ref{eq:ratedependence} yielded
$R_c(L)=16220\times\sqrt{L}~\rm{pairs}~s^{-1}mm^{1/2}$ with
excellent agreement to the square root dependence on $L$. The photon
bandwidths $\Delta\lambda_{s,i}$ were determined with a
single-photon Michelson interferometer and in Fig.
\ref{fig:lscaling} directly compared to the single-mode bandwidths
predicted by theory (Eq. \ref{eq:bandwidth}).

\section{Entangled Photon Source Setup} \label{sec:setup}
\begin{figure}[!bthp]
\centering\includegraphics[width=6.5cm]{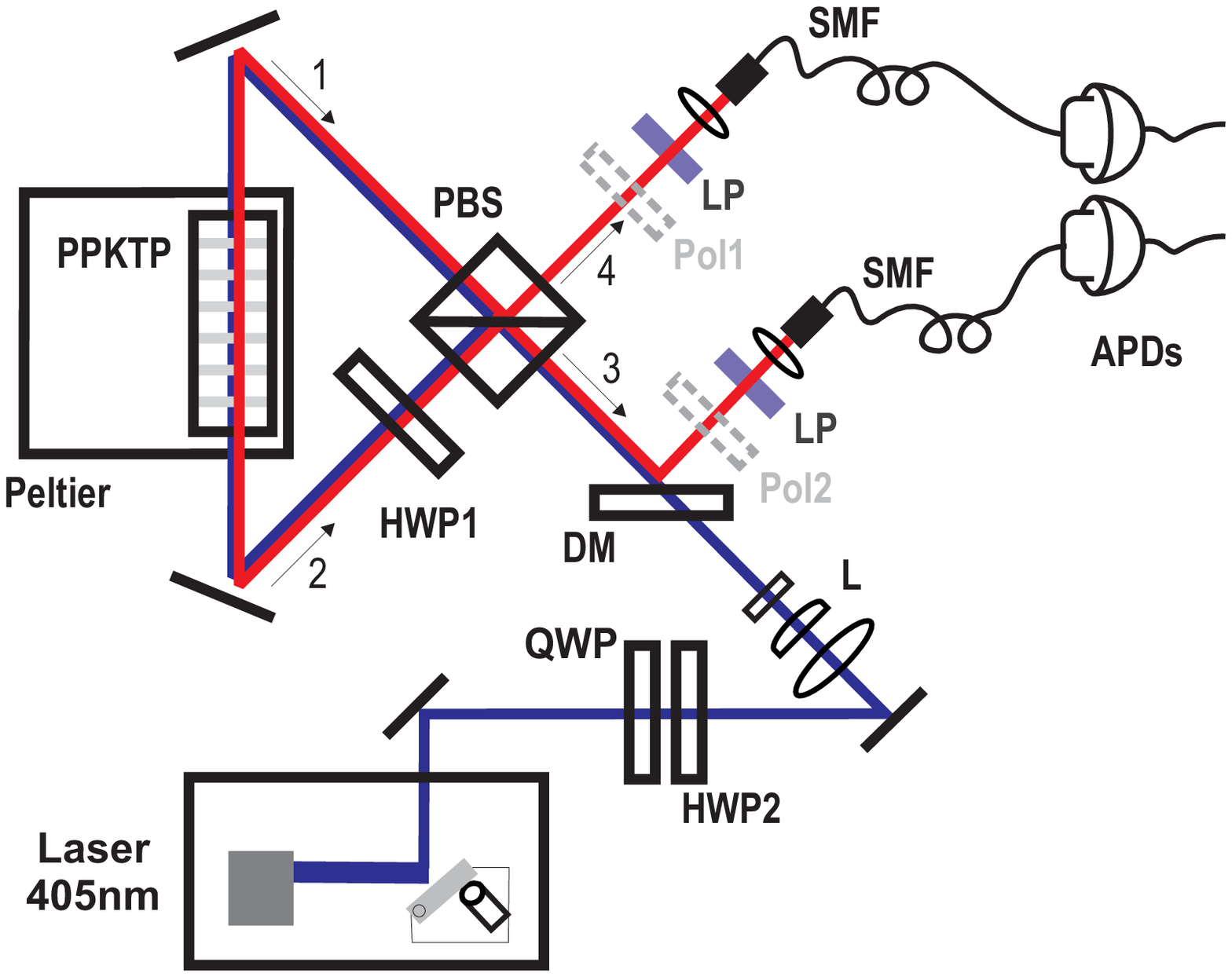} \caption{Scheme
of the fiber-coupled entangled photon source. A 405nm
\textit{Littrow} diode laser whose polarization and phase are set by
a quarter- (QWP) and a half-wave plate (HWP2) is focused via a
combination of one spherical and two cylindrical lenses (L) into a
PPKTP crystal inside a polarization Sagnac loop built up by
dual-wavelength polarizing beamsplitter (PBS) a dual-wavelength
half-wave plate (HWP1) (anti-reflection coated for $405$ and
$810$~nm) and two laser mirrors. The downconversion modes are
coupled into single-mode fibers (SMF). Remaining stray laser light
is blocked by two RG715 longpass color glass filters (LP).
Polarizers (Pol1, Pol2) can be inserted to characterize the produced
photon pairs.} \label{fig:setup1}
\end{figure}
Once we had found the optimal focusing conditions for the PPKTP
crystals and ascertained that the use of long crystals indeed
yielded more photon pairs and narrower bandwidth, the entangled
photon source was set up. The source is pumped by a \textit{Toptica}
$405$~nm grating-stabilized laser diode, focused to a circular spot
of waist $w_o=27~\mu m$ at the center of the crystal. The
polarization state of the laser beam is controlled by a combination
of a halfwave- (HWP2) and a quarter-wave plate (QWP). The $25$~mm
PPKTP crystal is mounted in a crystal oven, at the center of a
polarization Sagnac interferometer (PSI) which consists of a dual
wavelength polarizing beamsplitter cube (PBS), a dual wavelength
half-wave plate (HWP1) oriented at $\frac{\pi}{4}$ and two laser
mirrors (see Fig. \ref{fig:setup1}). The vertical component of the
laser beam is reflected at the PBS and rotated to horizontal
orientation by HWP1. It creates pairs of photons with orthogonal
polarizations $\ket{H_{s}}_1$ and $\ket{V_{i}}_1$, the number
subscripts denoting the spatial mode of the photons, which propagate
in clockwise direction. The horizontal component of the beam is
transmitted at the PBS and likewise creates pairs $\ket{H_{s}}_2$
and $\ket{V_{i}}_2$ propagating in counterclockwise direction. At
HWP1, $\ket{H_{s}}_2$ and $\ket{V_{i}}_2$ are rotated by
$\frac{\pi}{2}$ into $\ket{V_{s}}_2$ and $\ket{H_{i}}_2$. The
counterpropagating pairs in modes $1$ and $2$ are then combined at
the PBS, where again horizontal photons are transmitted and vertical
photons reflected such that
$\ket{H_{s}}_1,\ket{V_{i}}_1\rightarrow\ket{H_{s}}_3,\ket{V_{i}}_4$
and
$\ket{V_{s}}_2,\ket{H_{i}}_2\rightarrow\ket{V_{s}}_3,\ket{H_{i}}_4$.
The two-photon state emerging in modes $3$ and $4$ is thus
$\ket{H_{s}}_3\ket{V_{i}}_4+e^{i\phi}\ket{V_{s}}_3\ket{H_{i}}_4$.
The phase $\phi$ is set via appropriate adjustment of HWP2 and QWP
\cite{kim06}. Output mode $3$ is identical to the pump input mode,
therefore the downconversion photons have to be separated from the
pump by a dichroic mirror (DM). Subsequently, the photons in modes
$3$ and $4$ are coupled into single-mode optical fibers whose waists
are matched via $f=18.4$~mm aspheric lenses to the pump waist in the
crystal according to the focus parameters given in section
\ref{sec:focussing}. Since the single-mode fibers are spatial mode
filters for the downconversion light, there is no need for
additional lossy interference filters as in \cite{kim06}.

Note that, due to birefringence, orthogonally polarized photons in
the crystal travel at different group velocities, causing the
$\ket{H}$ photon to leave the crystal before the $\ket{V}$ photon.
This longitudinal walkoff renders the photons partly distinguishable
and leads to a decrease in the degree of entanglement. This effect,
which is inherent to all Type-II downconversion schemes, is
compensated by flipping the two photons in the counterclockwise arm
of the PSI by $\frac{\pi}{2}$. The longitudinal walkoff of pair $2$
is inverted with respect to pair $1$ and while there is still always
a non-zero delay between $\ket{H}$ and $\ket{V}$ photons in modes
$3$ and $4$, the arrival time of the photons contains no information
about the respective photon polarization anymore. The efficiency of
this compensation scheme depends only on the precision and
orientation of HWP1 and is thus more effective than in other schemes
\cite{kwiat95}.

\section{Entangled photon source performance and tunability}
\label{sec:results} The use of a $25~$mm long PPKTP crystal in the
perfectly compensated PSI scheme and the optimum choice of focus
parameters results in unprecedented photon yields, spectral
brightness and a very high degree of entanglement. Using a
\textit{PerkinElmer} SPCM-AQ4C single photon detector array with a
quantum efficiency of $\sim40\%$ and a home-built FPGA coincidence
counter with a coincidence time window of $4.4$~ns, we detected
$82000$ photon pairs/s at $1$~mW of pump power.
\begin{figure}[!bhtp]
\begin{center}
$\begin{array}{c@{\hspace{0.5in}}c} \multicolumn{1}{l}{\mbox{\bf
(a)}} & \multicolumn{1}{l}{\mbox{\bf (b)}} \\ [-0.53cm]
\includegraphics[width=3.5cm]{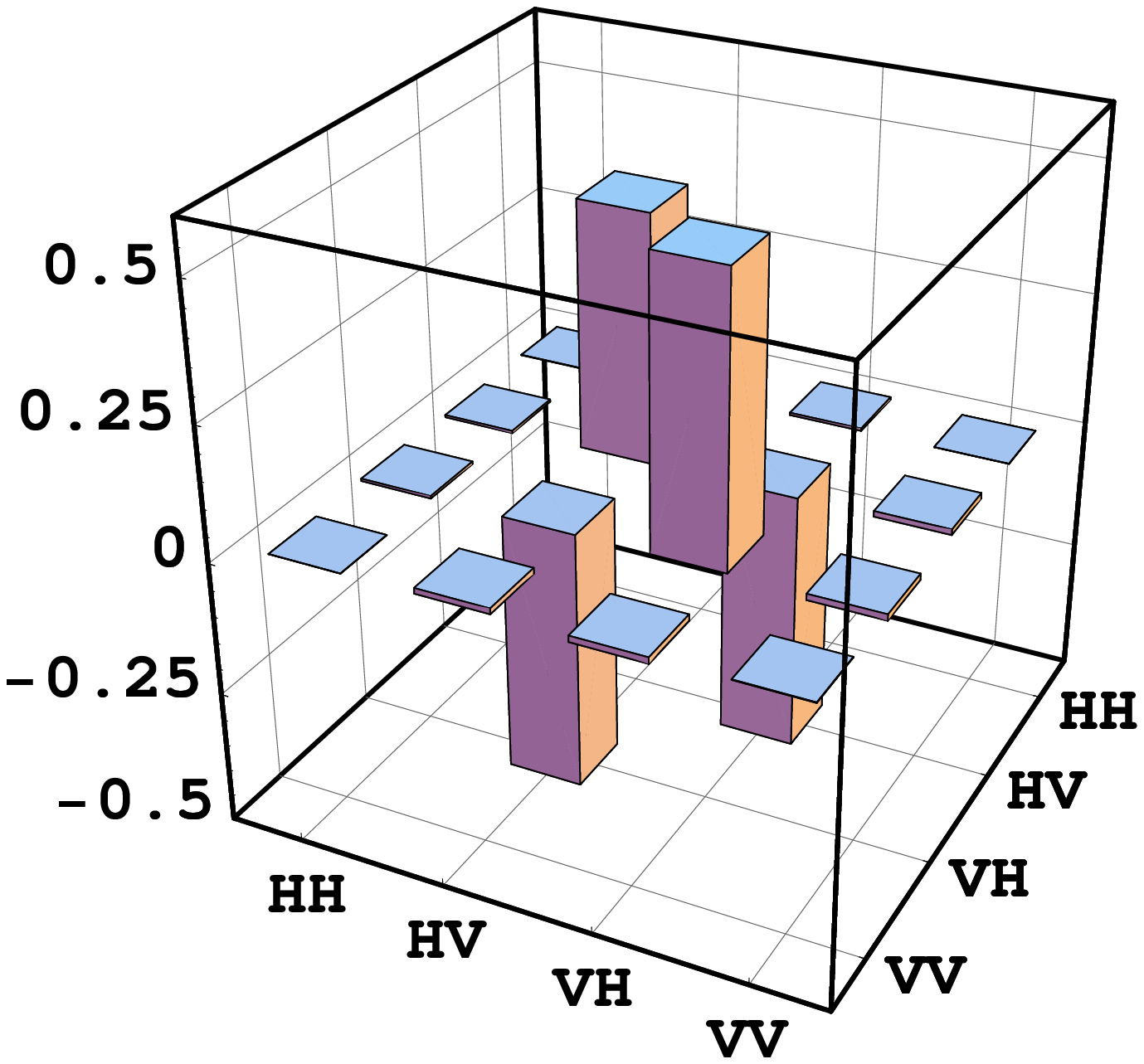} &
\includegraphics[width=3.5cm]{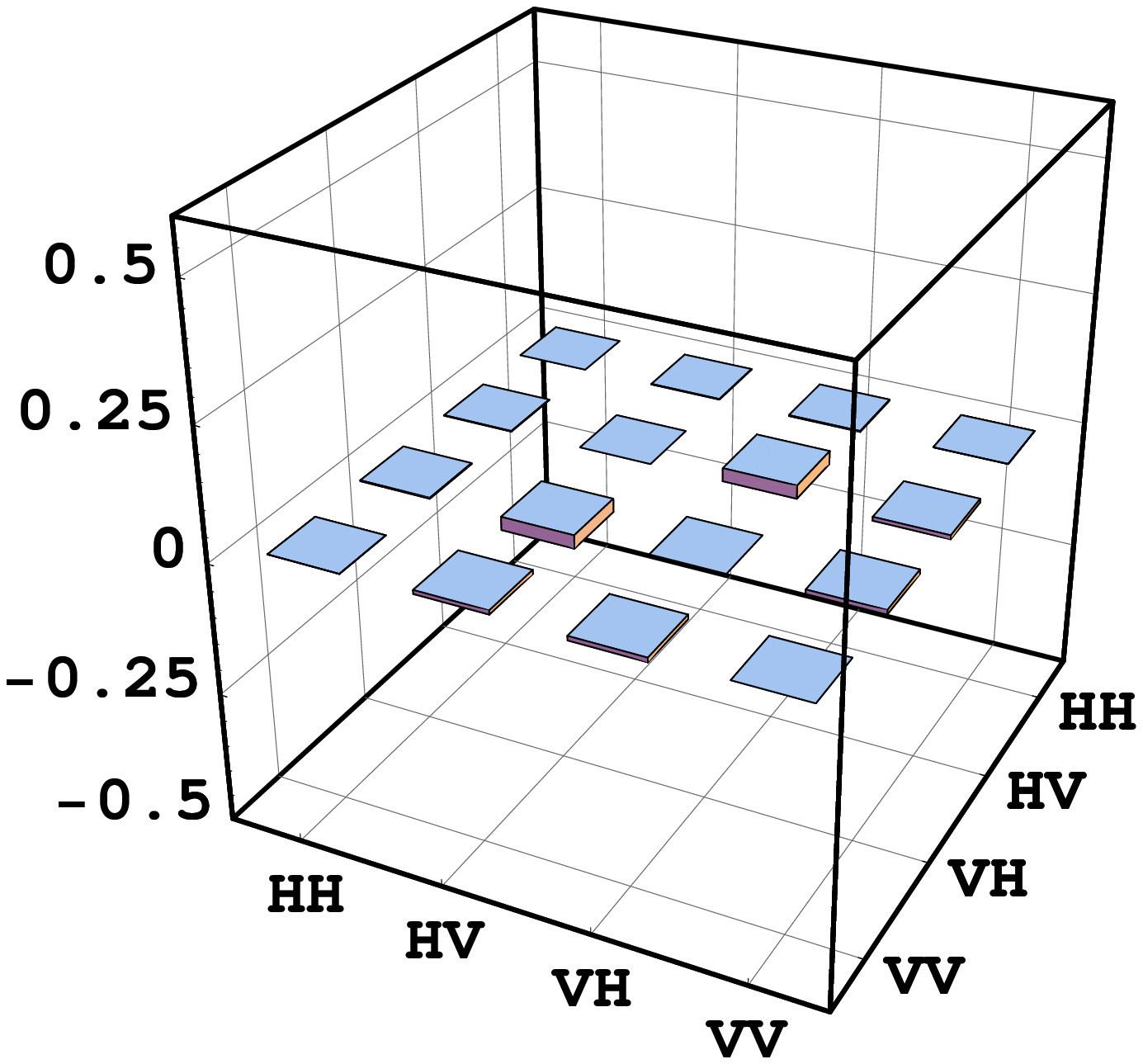}
\end{array}$
\end{center}
\caption{(color online) Tomography of the two-photon quantum state.
The real (a) and imaginary (b) part of the density matrix was
reconstructed from 36 linearly independent coincidence measurements.
The resulting fidelity $F$ to the $\ket{\psi^{-}}$ state was
$F=0.9959\pm0.0001$ and the Tangle $T=0.9875\pm0.0003$. Within the
resolution of this representation the experimental graph is in
excellent agreement to the theoretic expectation.} \label{fig:tomo}
\end{figure}

The overall transmission of the optical components in the setup
including the single-mode fibers, which were not AR-coated, and the
$100:1$ extinction ratio of the PBS, averaged over the optical paths
for output mode 3 and 4 amounts to $84.6\%$. Combined with the
detector efficiency, the maximal achievable coupling ratio
$\eta_c^0$ is therefore $33.8\%$. The actually observed $\eta_c$ was
$28.5\%$. This shows that in our setup the mode overlap
$\eta_c/\eta_c^0$, i.e. the probability that one photon of a pair is
collected given that its partner photon is in its respective
collection mode, is $84.3\%$. When set to the optimal focusing
conditions for $\eta_{c~\rm{max}}$, the photon pair yield of the
source decreases by $20\%$ to $65000$ pairs/s while the mode overlap
increases to $95\%$, to our knowledge the highest value reported so
far.

The photons have a measured $FWHM$ bandwidth of $0.3$~nm (see Fig.
\ref{fig:lscaling}), the spectral brightness of the source is
therefore $273000$~pairs~(s~mW~nm)$^{-1}$. The visibility of the
two-photon quantum state measured for degenerate photons
($\lambda_{s,i}=810$~nm) in the
$\ket{\pm}=\frac{1}{\sqrt{2}}(\ket{H}\pm\ket{V})$ basis was
$99.5\%$. The produced entangled state was further characterized
using full quantum state tomography \cite{james01}. The fidelity
$F=\bra{\psi^-}\rho\ket{\psi^-}$ of the two-photon density matrix
$\rho$ with the maximally entangled state
$\ket{\psi^-}=\frac{1}{\sqrt{2}}(\ket{H}\ket{V}-\ket{V}\ket{H})$ was
$F=0.9959\pm0.0001$ and the tangle, a measure for the degree of
entanglement as defined in \cite{coffman00} was $T=0.9875\pm0.0003$.
The measurements for visibility and quantum state tomography were
performed at $0.25$~mW of pump power and averaged over $10$ seconds.
The statistical standard deviations of these results were estimated
by performing a 100 run Monte Carlo simulation of the state
tomography analysis, with Poissonian noise added to the count
statistics in each run~\cite{james01}. Figure \ref{fig:tomo} shows
the reconstructed real and imaginary parts of the density matrix
$\rho$ of the produced states. After subtraction of accidental
coincidences caused by detector darkcounts, fidelity and tangle
increase to $F=0.9978\pm0.0001$ and $T=0.9940\pm0.0001$,
respectively. The state is thus very close to a pure entangled
state. Indeed in the $H/V$ basis we measured polarization contrasts
of up to 9500:1 (accidentals subtracted), a regime where visibility
measurements are limited by the extinction ratio of the polarizers,
which was found to be $10000:1$.

\begin{figure}[!bthp]
\centering\includegraphics[width=10cm]{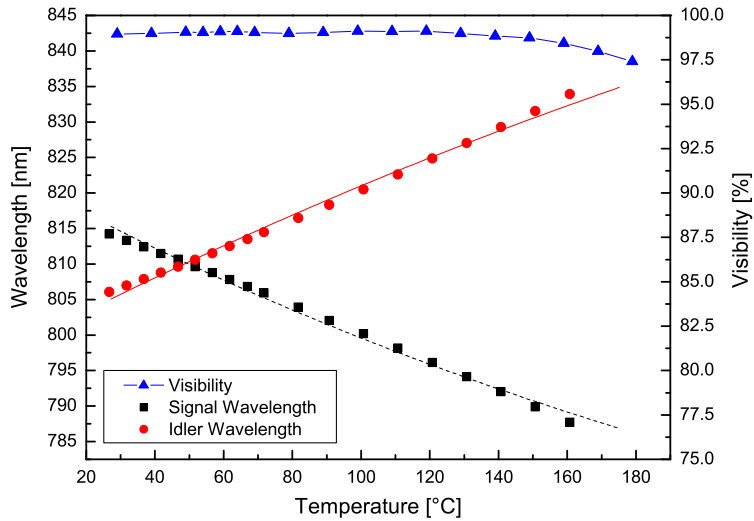}
\caption{Signal (black rectangles) and idler (red circles)
wavelengths and 2-photon visibility (blue triangles) as a function
of crystal temperature. The wavelengths were measured with a
scanning single photon spectrometer with a resolution of $0.2$~nm.
Our crystal oven allows us to reach a wavelength spacing of more
than $50$~nm.} \label{fig:visibility}
\end{figure}

The exact crystal temperature for wavelength degeneracy,
$49.2~\symbol{23}C$, was calculated from theory via Eq.
\ref{eq:momentum}. We experimentally confirmed this temperature with
a single-photon spectrometer, measuring $\lambda_i$ and $\lambda_s$
for temperatures from $25~\symbol{23}C$ to $60~\symbol{23}C$ (see
Fig. \ref{fig:visibility}). The most striking feature of the setup
is that the visibility of the entangled state stays above $99\%$ in
a large wavelength range (Fig. \ref{fig:visibility}). It starts to
decrease slightly for wavelength shifts of more than $\pm18~$nm,
reaching $97.5\%$ at $810\pm26~$nm at $\sim180\symbol{23}C$. The
most likely cause for this decrease is that all optical components
in the setup were specified and coated for $810~$nm and act as
partial polarizers at substantially higher or lower wavelengths. The
refractive indices needed to calculate theoretical values for
$\lambda_i$ and $\lambda_s$, according to Eq. \ref{eq:momentum},
were obtained via the Sellmeier and temperature-dependent Sellmeier
equations for KTP from \cite{kato03} and \cite{emanueli03}. The
thermal expansion coefficient for KTP which affects the crystal
poling period $\Lambda$ also stems from \cite{emanueli03}. Note that
these equations are of an empirical nature and do not exactly
reproduce the actual refractive indices for the whole spectral and
temperature range. In particular, the temperature-dependent
extensions for the Sellmeier equations from \cite{emanueli03} do not
extend to wavelengths lower than $532~$nm. In order to reproduce the
crystal temperature for degenerate downconversion and to obtain an
accurate wavelength fit over the whole temperature range, we had to
adopt $\partial n_y/\partial t\mid_{405nm}=28\times10^{-6}$,
reported in \cite{fiorentino05}.

\section{Conclusion}
We have demonstrated a fiber-coupled, wavelength-tunable source of
narrowband, polarization-entangled photons. By comprehensive
optimization of focusing conditions and the use of a $25~$mm long
crystal, we achieved a spectral brightness of
$273000$~(s~mW~nm)$^{-1}$, a factor of $28$ higher than comparable
previous setups. The entangled state has a tangle $T=0.987$ and the
coupling ratio is $28.5\%$. We demonstrated wavelength tuning of
entangled photons in a range of $\pm26$ nm around degeneracy with
virtually no decrease in entanglement. These properties, combined
with the intrinsic phase stability of the Sagnac-type setup, the
compactness, and the ease of use, make this setup a perfect tool for
quantum communication experiments both in the laboratory and in the
field.

\setcounter{secnumdepth}{0}
\section{Acknowledgements}
Thanks are due to Hannes Huebel for valuable help in starting up the
project, to Bibiane Blauensteiner for the single photon spectrometer
and to Robert Prevedel and Nathan Langford for helpful discussions.
We acknowledge support by the Austrian Science Foundation (FWF),
project number SFB1520, the Austrian Space Agency within the ASAP
program, the EC funded program QAP, the DTO-funded U.S. Army
Research Office and the City of Vienna.

\begin{thebibliography}{99}
\bibitem{ekert1991}%
A.~K.~Ekert, ``Quantum cryptography based on Bell's theorem,'' \prl
{\bf 67}, 661-663 (1991).

\bibitem{jennewein00}
T.~Jennewein, C.~Simon, G.~Weihs, H.~Weinurter, and A.~Zeilinger,
``Quantum cryptography with entangled photons,'' \prl {\bf 84},
4729-4732 (2000).

\bibitem{bennett93}
C.~H.~Bennett, G.~Brassard, C.~Cr{\'e}peau, R.~Jozsa, A.~Peres, and
W.~K.~Wootters, ``Teleporting an unknown quantum state via dual
classical and Einstein-Podolsky-Rosen channels,'' \prl {\bf 70},
1895-1899 (1993).

\bibitem{bouwmeester97}%
D.~Bouwmeester, J.~W.~Pan, K.~Mattle, M.~Eibl, H.~Weinfurter, and
A.~Zeilinger, ``Experimental quantum teleportation,'' \nat {\bf
390}, 575-579 (1997).

\bibitem{knill2001}%
E.~Knill, R.~Laflamme, and G.~J.~Milburn, ``A scheme for efficient
quantum computation with linear optics,'' \nat {\bf 409}, 46-52
(2001).

\bibitem{kwiat95}%
P.~G.~Kwiat, K.~Mattle, H.~Weinfurter, A.~Zeilinger,
A.~V.~Sergienko, and Y.~Shih, ``New high-intensity source of
polarization-entangled photon pairs,'' \prl {\bf 75}, 4337-4341
(1995).

\bibitem{pelton04}%
M.~Pelton, P.~Marsden, D.~Ljunggren, M.~Tengner, A.~Karlsson,
A.~Fragemann, C.~Canalias, and F.~Laurell, ``Bright,
single-spatial-mode source of frequency non-degenerate,
polarization-entangled photon pairs using periodically poled KTP,''
\opex {\bf 12}, 3573-3580 (2004).

\bibitem{konig05}
F.~K{\"o}nig, E.~J.~Mason, F.~N.~C.~Wong, and M.~A.~Albota,
``Efficient and spectrally bright source of polarization-entangled
photons,'' \pra {\bf 71}, 033805 (2005).

\bibitem{hubel07}%
H.~H\"{u}bel, M.~R.~Vanner, T.~Lederer, B.~Blauensteiner,
T.~Lor\"{u}nser, A.~Poppe, and A.~Zeilinger, ``High-fidelity
transmission of polarization encoded qubits from an entangled source
over 100 km of fiber,'' \opex {\bf 15}, 7853-7862 (2007).

\bibitem{kuklewicz04}
C.~E.~Kuklewicz, M.~Fiorentino, G.~Messin, F.~N.~C.~Wong, and
J.~H.~Shapiro, ``High-flux source of polarization-entangled photons
from a periodically poled KTiOPO$_4$ parametric down-converter,''
\pra {\bf 69}, 013807 (2004).

\bibitem{kwiat94}
P.~G.~Kwiat, P.~H.~Eberhard, A.~M.~Steinberg, and R.~Y.~Chiao,
``Proposal for a loophole-free Bell inequality experiment,'' \pra
{\bf 49}, 3209-3220 (2004).

\bibitem{fiorentino04}
M.~Fiorentino, G.~Messin, C.~E.~Kuklewicz, F.~N.~C.~Wong, and
J.~H.~Shapiro, ``Generation of ultrabright tunable polarization
entanglement without spatial, spectral, or temporal constraints,''
\pra {\bf 69}, 041801 (2004).

\bibitem{shi04}
B.~S.~Shi and A.~Tomita, ``Generation of a pulsed polarization
entangled photon pair using a Sagnac interferometer,'' \pra {\bf
69}, 013803 (2004).

\bibitem{kim06}%
T.~Kim, M.~Fiorentino, and F.~N.~C.~Wong, ``Phase-stable source of
polarization-entangled photons using a polarization Sagnac
interferometer,'' \pra {\bf 73}, 012316 (2006).

\bibitem{wong06}
F.~N.~C.~Wong, J.~H.~Shapiro, and T.~Kim, ``Efficient generation of
polarization-entangled photons in a nonlinear crystal,'' Laser
Physics, {\bf 16}, 1517-1524, (2006).

\bibitem{fan07}%
J.~Fan and A.~Migdall, ``A broadband high spectral brightness
fiber-based two-photon source,'' \opex {\bf 15}, 2915-2920 (2007).

\bibitem{liang07}
C.~Liang, K.~F.~Lee, M.~Medic, P.~Kumar, R.~H.~Hadfield, and
S.~W.~Nam, ``Characterization of fiber-generated entangled photon
pairs with superconducting single-photon detectors,'' \opex {\bf
15}, 1322-1327, (2007).

\bibitem{spillane07}%
S.~M.~Spillane, M.~Fiorentino, and R.~G.~Beausoleil, ``Spontaneous
parametric down conversion in a nanophotonic waveguide,'' \opex
{\bf{15}}, 8770-8780, (2007).

\bibitem{groeblacher07}%
S.~Gr{\"o}blacher, T.~Paterek, R.~Kaltenbaek, C.~Brukner,
M.~Zukowski, M.~Aspelmeyer, and A.~Zeilinger, ``An experimental test
of non-local realism,'' \nat {\bf 446}, 871-875, (2007).

\bibitem{paterek07}
T.~Paterek, A.~Fedrizzi, S.~Gr{\"o}blacher, T.~Jennewein,
M.~Zukowski, M.~Aspelmeyer, and A.~Zeilinger, ``Experimental test of
non-local realistic theories without the rotational symmetry
assumption,'' arXiv:0708.0813v1 [quant-ph]

\bibitem{resch05}%
K.~J.~Resch, M.~Lindenthal, B.~Blauensteiner, H.~R.~Boehm,
A.~Fedrizzi, C.~Kurtsiefer, A.~Poppe, T.~Schmitt-Manderbach,
M.~Taraba, R.~Ursin, P.~Walther, H.~Weier, H.~Weinfurter, and A.
Zeilinger, ``Distributing entanglement and single photons through an
intra-city, free-space quantum channel,'' \opex {\bf 13}, 1, 202-209
(2005).

\bibitem{peng05}%
C.~Z.~Peng, T.~Yang, X.~H.~Bao, J.~Zhang, X.~M.~Jin, F.~Y.~Feng,
B.~Yang, J.~Yang, J.~Yin, Q.~Zhang, N.~Li, B.~L.~Tian, and
J.~W.~Pan, ``Experimental free-space distribution of entangled
photon pairs over 13km: towards satellite-based global quantum
communication,'' \prl {\bf 94}, 150501 (2005).

\bibitem{ursin07}%
R.~Ursin, F.~Tiefenbacher, T.~Schmitt-Manderbach, H.~Weier,
T.~Scheidl, M.~Lindenthal, B.~Blauensteiner, T.~Jennewein,
J.~Perdigues, P.~Trojek, B.~\"{O}mer, M.~F\"{u}rst, M.~Meyenburg,
J.~Rarity, Z.~Sodnik, C.~Barbieri, H.~Weinfurter, and A.~Zeilinger,
``Entanglement-based quantum communication over 144km,'' Nature
Physics {\bf 3}, 481-486 (2007).

\bibitem{matsukevich04}%
D.~N.~Matsukevich and A.~Kuzmich, ``Quantum state transfer between
matter and light,'' Science {\bf 306}, 663-666 (2004).

\bibitem{julsgaard04}%
B.~Julsgaard, J.~Sherson, I.~Cirac, J.~Fiurasek, and E.~S.~Polzik,
``Experimental demonstration of quantum memory for light,'' \nat
{\bf 432}, 482-486 (2004).

\bibitem{kurtsiefer01}%
C.~Kurtsiefer, M.~Oberparleitner, and H.~Weinfurter,
``High-efficiency entangled photon pair collection in type-II
parametric fluorescence,'' \pra {\bf 64}, 023802 (2001).

\bibitem{bovino03}%
F.~A.~Bovino, P.~Varisco, M.~A.~Colla, G.~Castagnoli,
G.~Di~Giuseppe, and A.~V.~Sergienko, ``Effective fiber-coupling of
entangled photons for quantum communication,'' \oc {\bf 227},
343-348 (2003).

\bibitem{ljunggren05}%
D.~Ljunggren and M.~Tengner, ``Optimal focusing for maximal
collection of entangled narrow-band photon pairs into single-mode
fibers,'' \pra {\bf 72}, 062301 (2005).

\bibitem{rubin94}%
M.~H.~Rubin, D.~N.~Klyshko, Y.~H.~Shi, and A.~V.~Sergienko, ``Theory
of two-photon entanglement in Type-II optical parametric
down-conversion,'' \pra {\bf 50}, 5122-5233 (1994).

\bibitem{james01}%
D.~F.~V.~James, P.~G.~Kwiat, W.~J.~Munro, and A.~G.~White,
``Measurement of qubits,'' \pra {\bf 64}, 052312 (2001).

\bibitem{coffman00}%
V.~Coffman, J.~Kundu, and W.~K.~Wootters, ``Distributed
entanglement,'' \pra {\bf 61}, 052306 (2000).

\bibitem{kato03}%
K.~Kato and E.~Takaoka, ``Sellmeier and thermo-optic formulas for
KTP,'' \ao {\bf 41}, 5040–5044 (2002).

\bibitem{emanueli03}%
S.~Emanueli and A.~Arie, ``Temperature-dependent dispersion
equations for KTiOPO4 and KTiOAsO4,'' \ao {\bf 42}, 6661-6665,
(2003).

\bibitem{fiorentino05}%
M.~Fiorentino, C.~E.~Kuklewicz, and F.~Wong, ``Source of
polarization entanglement in a single periodically poled KTiOPO$_4$
crystal with overlapping emission cones,'' \opex {\bf 13}, 1,
127-135 (2005).

\end{thebibliography}
\end{document}